\documentclass{frontiersSCNS} 

\usepackage{url,hyperref,lineno,microtype}
\usepackage[onehalfspacing]{setspace}

\newcommand{\aap}{    {\it Astron. Astrophys.}}

\newcommand{\apjs}{ {\it Astrophys. J. Suppl.}}
\newcommand{\apj}{    {\it Astrophys. J.}}
\newcommand{\apjl}{   {\it Astrophys. J. Lett.}}

\newcommand{\mnras}{  {\it Mon. Not. Roy. Astron. Soc.}}

\newcommand{\solphys}{{\it Solar Phys.}}

\def\keyFont{\fontsize{8}{11}\helveticabold }
\def\firstAuthorLast{Narang {et~al.}} 
\def\Authors{Nancy Narang\,$^{1,*}$, Vaibhav Pant\,$^{2}$, Dipankar Banerjee\,$^{1}$ and Tom Van Doorsselaere\,$^{2}$}
\def\Address{$^{1}$Indian Institute of Astrophysics, Koramangala 2nd Block, Bangalore, India \\
$^{2}$Centre for mathematical Plasma Astrophysics, Mathematics Department, KU Leuven, Belgium}
\def\corrAuthor{Nancy Narang, Indian Institute of Astrophysics, Koramangala 2nd Block, Bangalore - 560034, India}

\def\corrEmail{nancy@iiap.res.in}

\begin{document}
\onecolumn
\firstpage{1}

\title[High-frequency dynamics of moss]{High-frequency dynamics of active region moss as observed by IRIS} 

\author[\firstAuthorLast ]{\Authors} 
\address{} 
\correspondance{} 

\extraAuth{}

\maketitle

\begin{abstract}

\section{}
The high temporal, spatial and spectral resolution of Interface Region Imaging Spectrograph (IRIS) has provided new insights into the understanding of different small-scale processes occurring at the chromospheric and transition region (TR) heights. We study the dynamics of high-frequency oscillations of active region (AR 2376) moss as recorded by simultaneous imaging and spectral data of IRIS. Wavelet transformation, power maps generated from slit-jaw images in the Si\,IV\,1400\,\AA~passband, and sit-and-stare spectroscopic observations of the Si\,IV\,1403\,\AA~spectral line reveal the presence of high-frequency oscillations with $\sim$1--2\,minute periods in the bright moss regions. The presence of such low periodicities is further confirmed by intrinsic mode functions (IMFs) as obtained by the empirical mode decomposition (EMD) technique. We find evidence of the presence of slow waves and reconnection-like events, and together they cause the high-frequency oscillations in the bright moss regions.

\tiny
 \keyFont{ \section{Keywords:} Sun, chromopshere, transition region, moss, oscillations, mhd wave} 
\end{abstract}

\section{Introduction}
\label{sec:intro}

Understanding the processes responsible for the heating of the upper atmosphere is the central problem in solar physics. Though highly debated (see reviews,~\citealp{klimchuk06, reale10, parnell12}), two widely accepted mechanisms for converting magnetic energy into thermal energy are impulsive heating by nano-flares \citep{parker88} and heating by dissipation of waves \citep{arregui15}. The heating processes are generally proposed to occur on small spatial and temporal scales, which were difficult to observe with the typical resolution of the previous instruments. In the very recent past, the advent of instruments with better temporal resolutions, several evidences of high-frequency oscillations of sub-minute periodicities have been reported to be present from the chromosphere~\citep{gupta15, shetye16, jafarzadeh17, ishikawa17} up to the corona~\citep{testa13, morton13, morton14, pant15, samanta16}) at sub-arcsec spatial scales. The small-scale quasi-periodic flows resulting from oscillatory magnetic reconnection as well as the presence of various Magnetohydrodynamic (MHD) waves produce such observed perturbations in imaging and spectroscopic observables. These periodic/quasi-periodic perturbations/oscillations observed at such finer scales in space and time can thus be regarded as the manifestations of the reoccurring dynamic heating processes present at similar spatial (sub-arcsec) and temporal (sub-minute) scales.

Various MHD waves  could  be present simultaneously along with quasi-periodic flows or their presence could entirely be non-concurrent. The plausible mechanism/s for their origin might also be directly coupled in some cases or completely independent in others. For instances,~\citet{gupta15} detected short-period variability (30--90\,s) within explosive events observed in TR by IRIS~\citep{depontieu14} and related them to repetitive magnetic reconnection events. On the other hand,~\citet{jafarzadeh17} observed high-frequency of periods 30--50\,s in Ca II H bright-points in the chromosphere using the SUNRISE Filter Imager (SuFI;~\citealp{gandorfer11}). They found the evidence of both compressible (sausage mode) and incompressible (kink mode) waves to be present in the magnetic bright-points.~\citet{shetye16} reported transverse oscillations and intensity variations ($\sim$ 20--60\,s) in chromospheric spicular structures using the CRisp Imaging SpectroPolarimeter (CRISP;~\citealp{scharmer08}) on the Swedish 1-m Solar Telescope. They argued that high-frequency helical kink motions are responsible  for transverse oscillations and compressive sausage modes to result in intensity variations. They further found evidence of mode coupling between compressive sausage and non-compressive kink modes and speculated the presence of other spicules and flows possibly acting as the external drivers for the mode-coupling.

Using the total solar eclipse observations of 11 July 2010~\citep{singh11},~\citet{samanta16} detected significant oscillations with periods $\sim$6--20\,s in coronal structures. They attributed these high-frequency oscillations as a mixture of different MHD waves and quasi-periodic flows. Using the High-resolution Coronal Imager (Hi-C;~\citet{kobayashi14}) data,~\citet{testa13} observed variability on time-scales of 15--30\,s to be present in the moss regions as observed in the upper TR, which they found to be mostly located  at the foot-points of coronal loops. They regarded such oscillations as the signatures of heating events associated with reconnection occurring in overlying hot coronal loops, \textit{i.e.}, impulsive nano-flares. More recently, from the Chromospheric Ly$\alpha$ SpectroPolarimeter (CLASP; ~\citealp{kano12}) observations,~\citet{ishikawa17} also reported short temporal variations in the solar chromosphere and TR emission of an active region with periodicities of $\sim$10--30\,s. They attributed these intensity variations to waves or jets from the lower layers instead of nano-flares.~\citet{morton13, morton14} analysed the same active region moss observations of Hi-C as by~\citet{testa13} and observed the presence of transverse oscillations with periodicities of 50--70\,s.~\citet{pant15} also studied the same region in Hi-C observations and detected quasi-periodic flows as well as transverse oscillations with short periodicities (30--60\,s) in braided structures of the moss. They indicated coupling between the sources of transverse oscillations and quasi-periodic flows, \textit{i.e.}, magnetic reconnection, such that they could be possibly driving each other.

In the present work, we concentrate on the high-frequency ($\sim$1--2\,minute) dynamics of active region (AR 2376) moss as observed by IRIS. IRIS have provided an unprecedented view of the solar chromosphere and transition region with high temporal, spatial and spectral resolution. The joint imaging and spectroscopic observations of IRIS at high cadence provide us with a unique opportunity to have a detailed analysis of different characteristics and mechanisms involved in the generation of high-frequency oscillations in TR moss regions.

\begin{figure}[htbp]
	\centering
	
	\includegraphics[width=15cm]{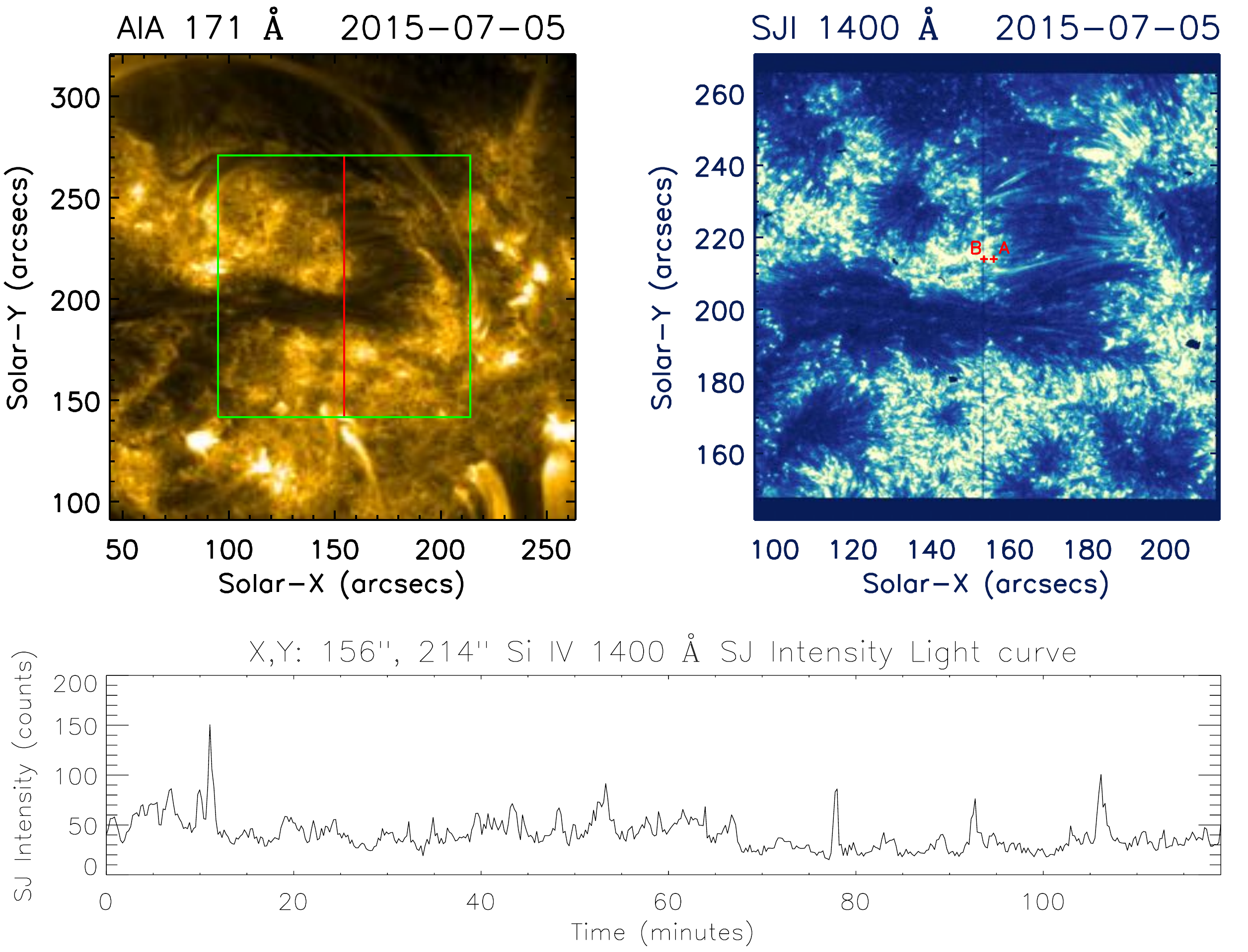}
	
	\caption{AIA 171\,\AA~ image marking the observed region by IRIS and SJI at a particular instance as observed by IRIS in the  Si\,IV\,1400\,\AA~passband. The bottom panel shows the SJ intesity light curve at location A marked in SJI FOV for the complete duration of the observation.}
	
	\label{fig1}
\end{figure}

\section{Details of the Observation}
\label{sec:obs}

IRIS observations of active region (AR 2376) moss, observed on 2015-07-05 from 05:16:15 UT to 07:16:23 UT is considered for the present analysis. Figure~\ref{fig1} shows the observation region on the solar disk, as outlined in the image taken in the 171\,\AA~ pass-band of AIA (Atmospheric Imaging Assembly;~\citealp{lemen12}) and slit-jaw image (SJI) in 1400\,\AA ~at a particular instance observed by the IRIS. The bottom panel shows a typical light-curve at a particular location A (marked in the full FOV above) in the moss region in SJ 1400\,\AA~intensity. The nature of the variation of intensity clearly reveals the presence of small amplitude quasi-periodic variations along with comparatively larger amplitude variations. 

Centred at $146^{\prime\prime},207^{\prime\prime}$, the imaging data (slit-jaw images or SJIs) have a field of view (FOV) of ~$119^{\prime\prime}\times119^{\prime\prime}$. The SJIs are taken with a cadence of 13 seconds and have spatial resolution $\approx0.33^{\prime\prime}$. The simultaneous large sit-and stare spectroscopic data has a cadence of 3.3 seconds with the slit-width of $0.35^{\prime\prime}$ and pixel size along the solar-Y axis to be $0.1664^{\prime\prime}$ with slit length of ~$119^{\prime\prime}$. Every observation in this data-set has an exposure time of 2 seconds. The high cadence of these data-sets provides us a unique opportunity to investigate the high-frequency dynamics in this region with high significance level.

We use IRIS SJIs centred at the  Si\,IV\,1400\,\AA~passband which samples emission from the transition region (TR). For spectral analysis, we concentrate on the  Si\,IV\,(1403\,\AA) line formed at $\log_{10}T\approx4.9K$ which is one of the prominent TR emission lines observed with the IRIS and is free from other line blends. For density diagnostics, we use the O\,IV\,(1401\,\AA) TR line along with Si\,IV\,(1403\,\AA)~\citep{keenan02,young18a}. The calibrated level 2 data of IRIS is used in the study. Dark current subtraction, flat-field correction, and geometrical correction have been taken into account in the level 2 data. We employ wavelet analysis~\citep{torrence98} and empirical mode decomposition (EMD;~\citealp{huang98}) techniques in order to detect and characterize the high-frequency oscillations in slit-jaw (SJ) intensity (section~\ref{sec:images}) and different spectral properties~\textit{i.e.}, total intensity, peak intensity, Doppler velocity, and Doppler width~(sections~\ref{sec:spectra}).




\begin{figure}[htbp]
	\centering
	\includegraphics[width=17cm]{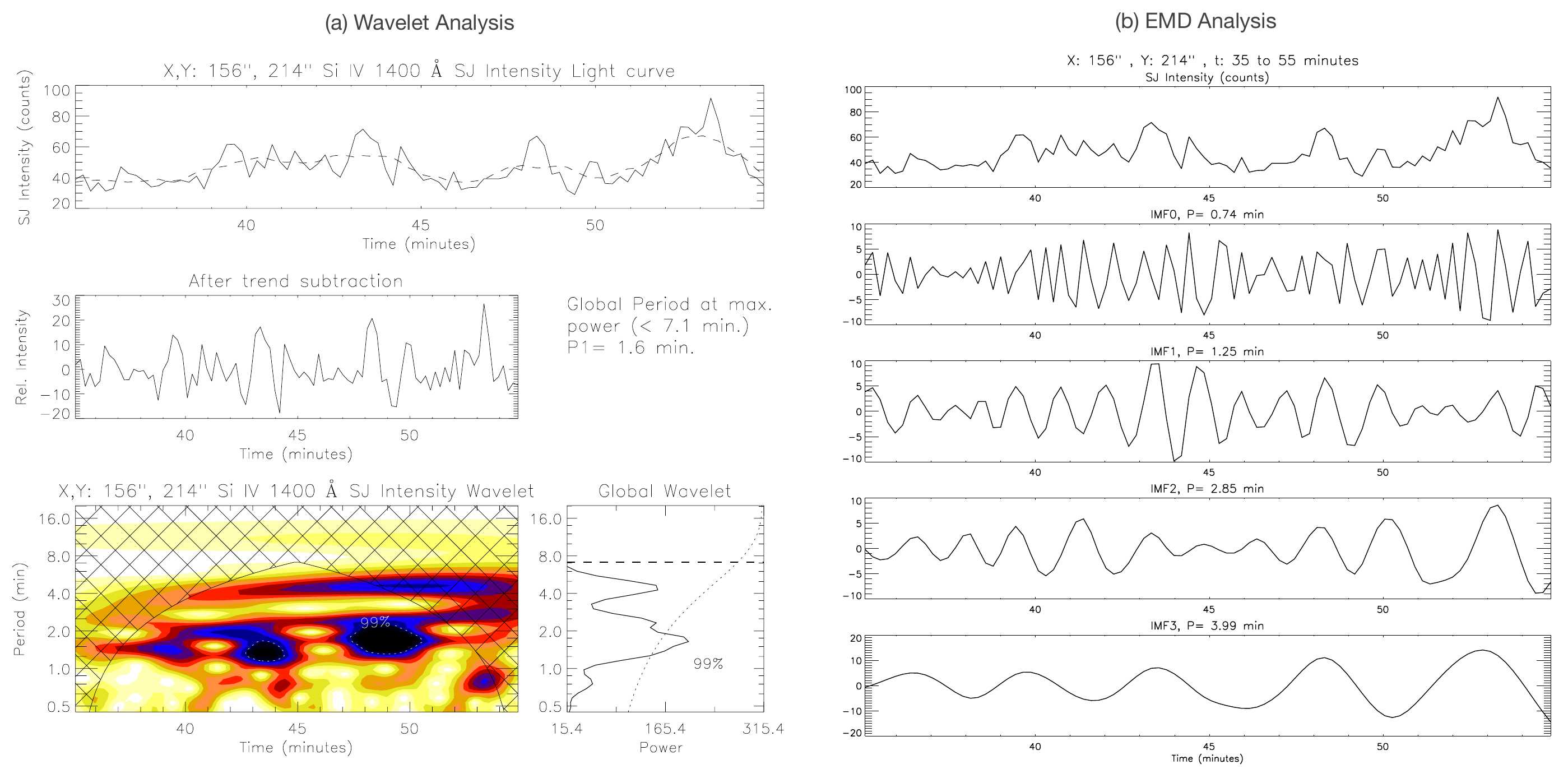}
	
	\caption{(a) Wavelet analysis, and (b) EMD analysis result for the Si\,IV\,1400\,\AA~SJ intensity variation with time from 35 to 55 minutes of the observation at location A. The details about the different panels are explained in the text (section~\ref{sec:images}).}
	
	\label{fig2}
\end{figure}

\section{Data-Analysis and Results}

\label{sec:result}

\subsection{Imaging Analysis from Si\,IV\,1400\,\AA~SJIs}

\label{sec:images}

Wavelet analysis is performed at each pixel location of SJ FOV to obtain the period of SJ intensity variability over the observed moss region. As shown in Figure~\ref{fig1}, a typical light curve corresponding to a single pixel location for the entire duration reveals presence of  quasi-periodic small and large amplitude intensity fluctuations.
Figure~\ref{fig2}\,(a) shows a representative example (selected at random) of wavelet analysis results corresponding to the pixel location marked as A in SJ FOV (Figure~\ref{fig1}) for a duration of 20\,minutes. It should be noted that in most of results we show the wavelet and EMD analysis corresponding to 20~minutes interval only so that the temporal variations in intensity can be studied more carefully, particularly as we are interested in the shorter periodicities. The top panel in Figure~\ref{fig2}\,(a) shows the variation of SJ intensity with time. The middle-panel shows the background (trend) subtracted intensity which is further used to obtain wavelet power spectrum (lower panels). The background (trend) is obtained by taking the 10-point running average of the intensity variation. The bottom left panel displays a wavelet power spectrum (color inverted) with 99\% significance levels and the bottom right panel displays a global wavelet power spectrum (wavelet power spectrum summed over time) with 99\% global significance.

\begin{figure}[htbp]
	\centering
	
	\includegraphics[width=16cm]{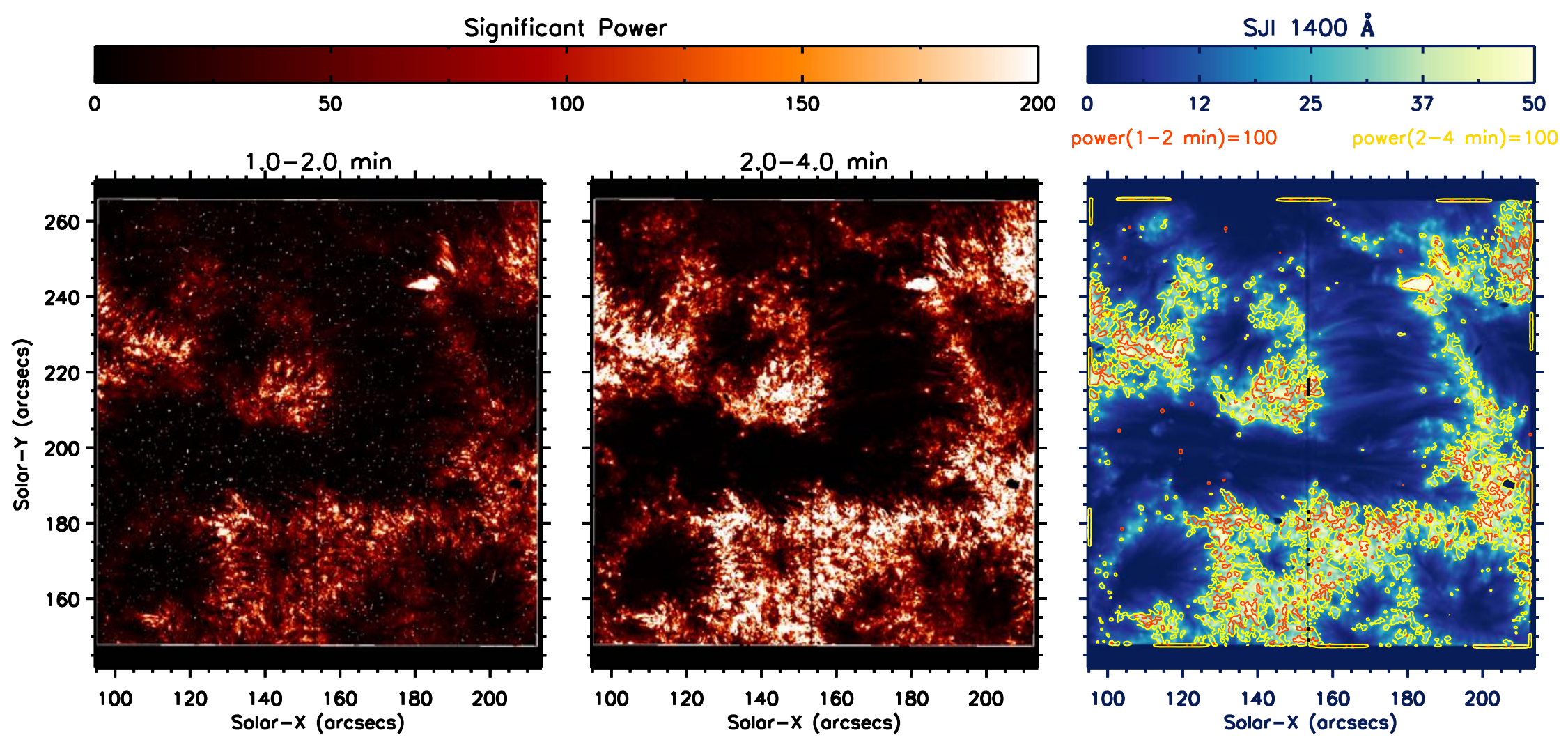}

	\caption{Power maps showing the significant power obtained from the Si\,IV\,1400\,\AA~SJ intensity variation in the period range of 1 to 2 minutes and 2 to 4 minutes. The rightmost panel shows the average SJ intensity image with contours of significant power in 1-2 minutes periods in red and 2-4 minutes in yellow, delineating the bright regions of the moss.}
	
	\label{fig3}
\end{figure}

The power spectra obtained reveals the presence of short-period variability in the SJ intensity light-curve, with a distinct power peak at a period of 1.6\,min. It is important to note that even without considering the background trend, we obtain a power peak at the same period in wavelet spectra but with low significance level. Empirical mode decomposition (EMD) is also employed at a few locations in the SJ FOV. Figure~\ref{fig2}\,(b) shows the different intrinsic mode functions (IMFs) obtained from EMD for the same SJ light-curve as shown in Figure~\ref{fig2}\,(a). Here only the first four IMFs are shown as the further IMFs contain the larger background trends. The dominant period (P) mentioned in the figure for each IMF is calculated using fast-Fourier transform (FFT). The period of the first four IMFs for the particular example shown in Figure~\ref{fig2}\,(b) are 0.74\,min, 1.25\,min, 2.85\,min and 3.99\,min. The EMD analysis reinforces the detection of the presence of short periodicities (1--2\,min) in the moss region as obtained by wavelet analysis. The presence of periodicities $<$\,1\,min can also be noted from the Figure~\ref{fig2}, though these are  below the significance level of 99\% as shown in wavelet power spectra. Such oscillations have very small amplitudes, are present even for shorter-duration and could be damping fast. Hence, the oscillations with periods $<$\,1\,min may carry smaller amount of energy and may not be so important as those with periods $>$\,1\,min which may be distributed over larger spatial and temporal extents.

To focus on the distribution of power as calculated from the wavelet method, we obtain the power maps of SJ intensity over the full FOV in 1--2\,min and 2--4\,min period intervals (Figure~\ref{fig3}) by considering the entire duration of the observation. The entire duration of the observations is chosen to understand the global dynamics of the active region moss. On comparison of power maps with the SJ images (Figure~\ref{fig3}) and AIA images (Figure~\ref{fig1}), it can be observed that the significant power of high-frequency (1--2\,min) as well as low-frequency (2--4\,min) oscillations is generally present only in bright regions of the moss. Figure~\ref{fig3} also shows the time-average SJI with the power contours of of 1--2\,min variability in red and 2--4\,min in yellow. The power contours enclose the locations with the value of significant power to be more than 100, in respective period range. The finer and smaller spatial extents of the contours at various locations over the field of view suggest that these oscillations possess high power in the localized regions within the bright moss. Moreover, the comparison of power between short (1--2\,min) and long (2--4\,min) periodicities, as showcased in Figure~\ref{fig3}, reveals that the power in 1--2\,min variability is, in general, less than that in 2--4\,min.


\begin{figure}[htbp]
	\centering
	
	\includegraphics[width=18cm]{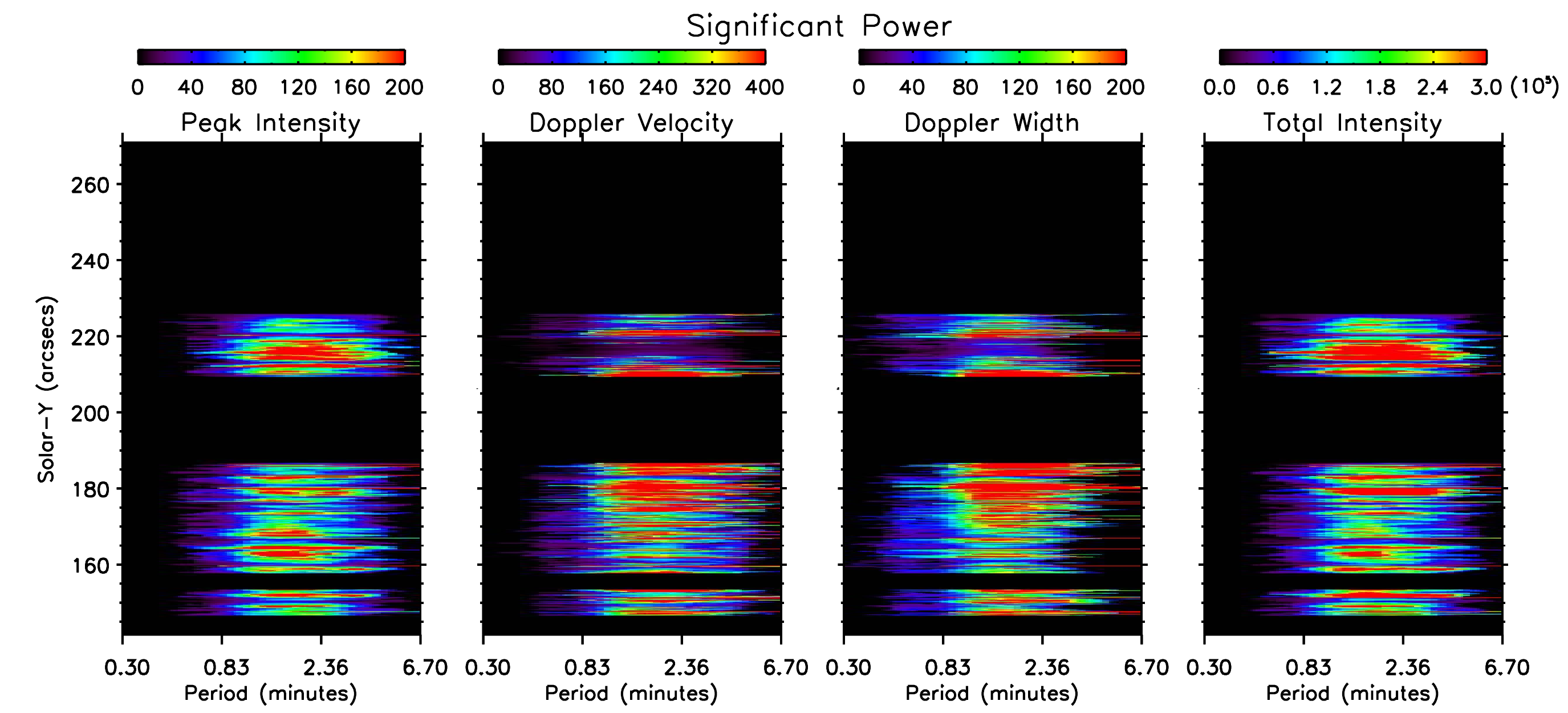}
	
	\caption{Power maps showing the significant power obtained from the variation of different spectral parameters of Si\,IV\,1403\,\AA~emission line in the period range of 0.3 to 6.7 minutes.}
	
	\label{fig7}
\end{figure}

\begin{figure}[htbp]
	\centering
	\includegraphics[width=17cm]{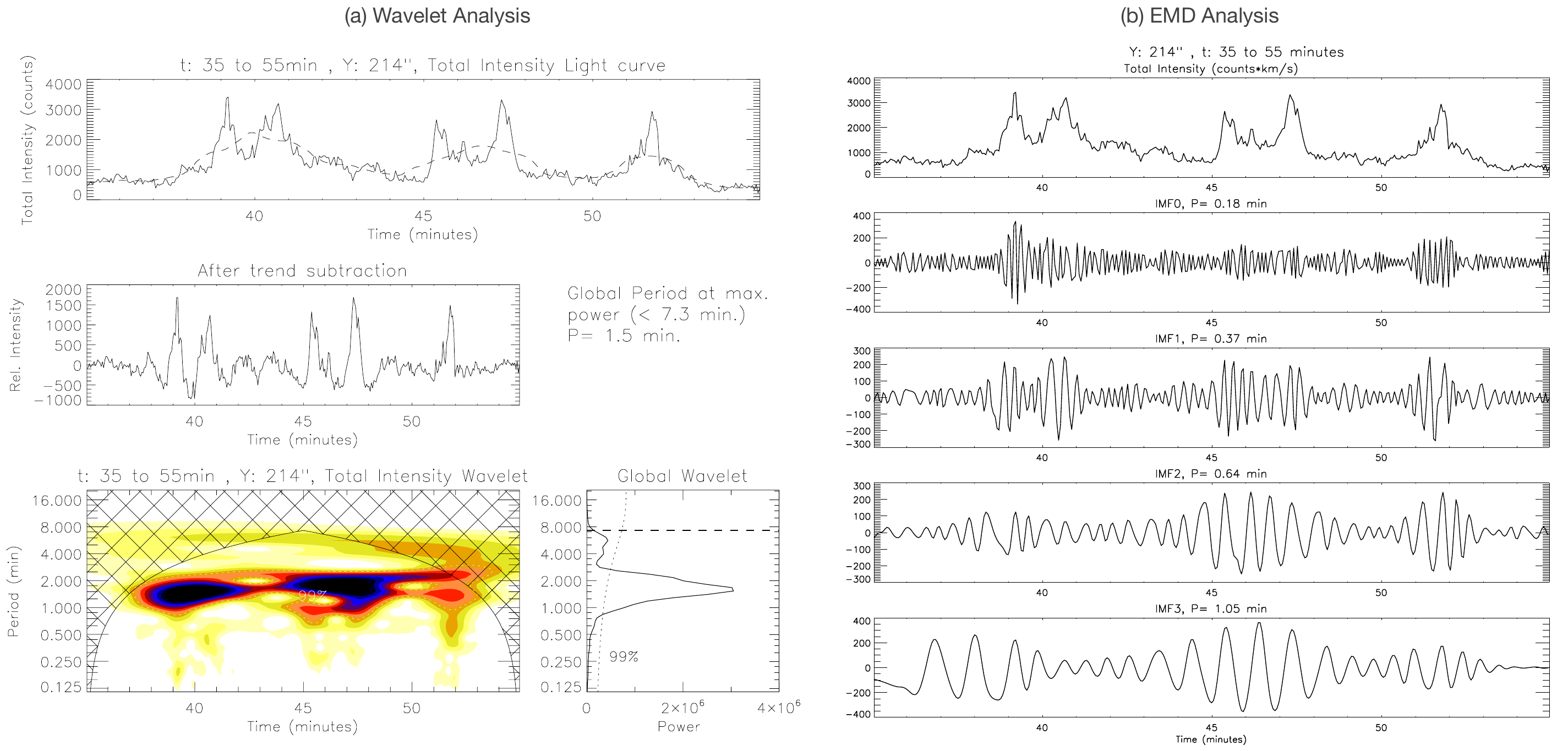}
	
	\caption{(a) Wavelet analysis, and (b) EMD analysis result for the Si\,IV\,1403\,\AA~total intensity variation with time from 35 to 55 minutes of the observation at location B along the slit.}
	
	\label{fig4}
\end{figure}

\begin{figure}[htbp]
	\centering
	\includegraphics[width=17cm]{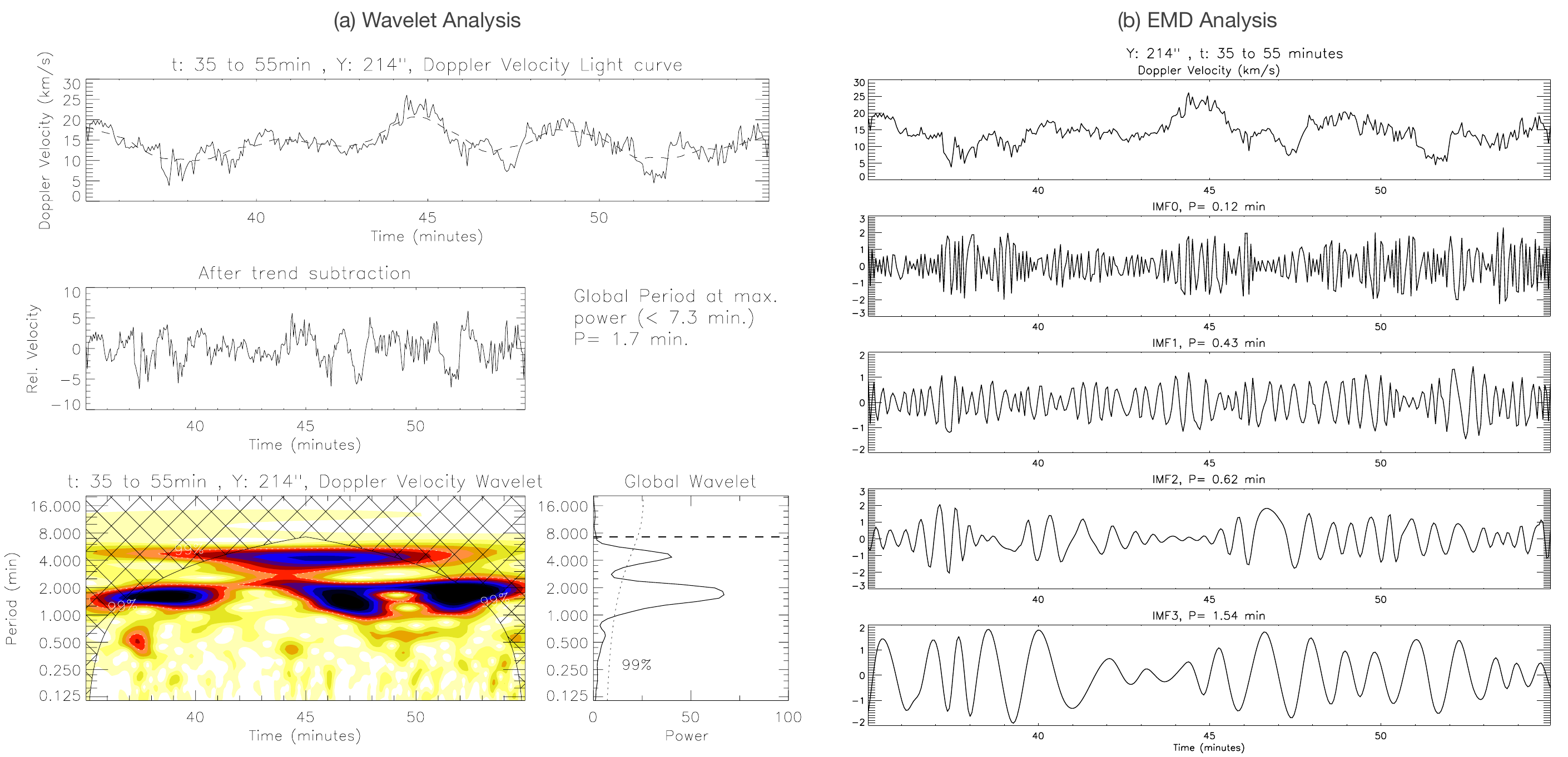}
	
	\caption{Same as figure~\ref{fig4} but for Doppler velocity}
	
	\label{fig5}
\end{figure}

\subsection{Spectral Analysis from Si\,IV\,1403\,\AA~emission line}

\label{sec:spectra}
To characterize periodicities present in the spectrograph data, we  produce power maps of the spectral parameters obtained by fitting a single Gaussian to the Si\,IV\,1403\,\AA~emission spectra using wavelet analysis. The analysis was performed over the entire duration of the observations.  At few instances, we interpolate the spectral parameters where a Gaussian fitting could not be performed due to poor signal to noise. The power maps clearly showcase the significant power along the slit, predominantly present in the period range of 0.83 to 2.36\,min corresponding to pixel locations of the bright moss regions (wherever the slit crosses the bright moss). This confirms the presence of short-period oscillations in Si\,IV\,1403\,\AA~spectra along with the Si\,IV\,1400\,\AA~SJ intensity (described in Section~\ref{sec:images}) in various locations of the moss region. 

\begin{figure}[htbp]
	\centering
	
	\includegraphics[width=18cm]{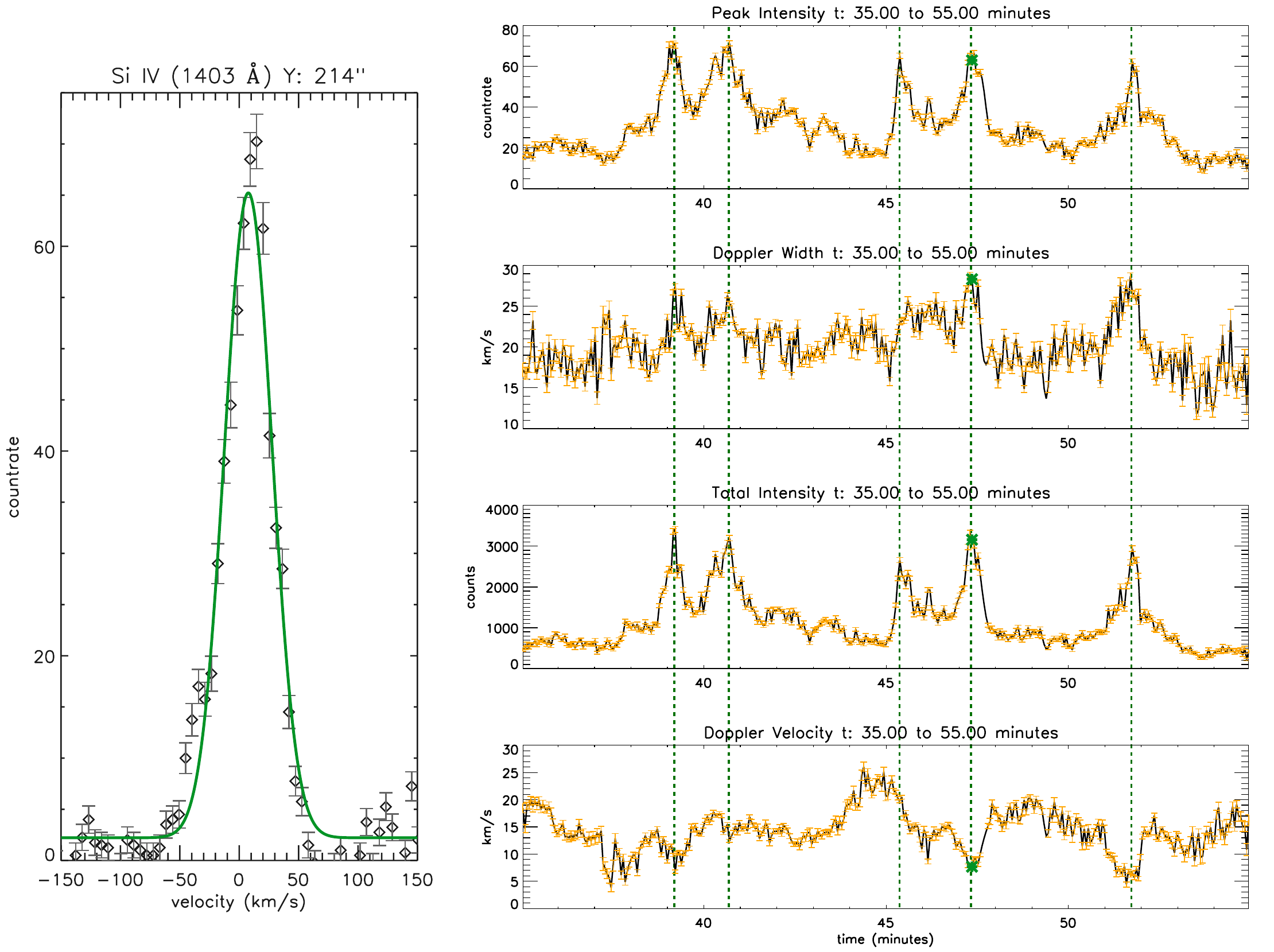}
	
	\caption{Left panel shows the observed spectral line profile (\textit{black} symbols with the observational errors) of Si\,IV\,1403\,\AA~emission line at location B, with the Gaussian fit (\textit{green} solid curve) for a particular instance. Panels on the right show the light-curves (in \textit{black}) with the fitting errors (in \textit{orange}) of different spectral parameters at B for a duration of 20\,minutes. The \textit{green} solid symbol over the light-curves marks the instant for which the spectral profile is shown in the left panel. The dotted lines shows some instances of the possible reconnection flows. An animation of this figure is available in the online version which shows the evolution of the spectral profile with time.}
	
	\label{fig6}
\end{figure}

Now we shift our focus to shorter time intervals where data gaps due to poor signal-to-noise are absent. This allow us to investigate the correlation between different spectral parameters using wavelet and EMD analysis. Further, taking the intervals of 20~minutes is sufficient because we are primarily interested  in shorter periods like, 1-2~minutes.  Figure~\ref{fig4}\,(a) shows the wavelet maps of total intensity variation for a duration of 20\,minutes at a particular location along the slit (marked as B in the SJ FOV in Figure~\ref{fig1}). Total intensity signifies the summed intensity over the wavelength range. Figure~\ref{fig5}\,(a) shows the wavelet maps of Doppler velocity at the same location B and same time-interval as shown for total intensity in Figure~\ref{fig4}. Note that the location B is very close to location A so that a comparison can be made with the periodicities found at location A using SJI. Moreover, the same time-interval is shown in Figures~\ref{fig2},\,\ref{fig4} and \ref{fig5} for better illustration. Figure~\ref{fig6} shows the variation of peak intensity, Doppler width, total intensity and Doppler velocity of Si\,IV\,1403\,\AA \, line at location B along with the spectral line-profile at a particular instance. The observational uncertainties are shown in the left panel over the observed line-profile. These errors are taken into account while fitting the Gaussian profile (\textit{green} solid curve). The fitting errors of the respective spectral parameters are shown in the adjacent light curves in \textit{orange}. It can be clearly observed that the errors in the spectral parameters are much less than the amplitude of oscillations. For instance, the average magnitude of error over the Doppler velocity light curve shown in Figure~\ref{fig5} and \ref{fig6} is 0.5 km/s, whereas the amplitude of oscillation of its IMFs (as shown in Figure~\ref{fig5}) is more than 1 km/s in most of the cases. The oscillations in the spectral parameters are well above the error values in general and thus significant. An animation of the Figure~\ref{fig6} is available in the online version which shows the evolution of the spectral line profile with time.

The background trends for the spectral parameter light curves (in Figure~\ref{fig4}\,and\,\ref{fig5}) are obtained by considering the 35-point running average of the light-curves. The dominant power peaks are observed to be present 1.5\,min for total intensity, 1.7\,min for peak intensity, 1.7\,min for Doppler velocity, and 1.5\,min for Doppler width in the respective power spectra. Here again, the presence of periodicities of $<$\,1\,min can be seen in the wavelet. It can be clearly observed that such oscillations are present for very short durations and thus of not much significance over the longer durations. Also, such short periodicities could be due to the presence of noise which is picked up by wavelet at higher-frequencies.

The EMD technique is applied over the spectral variations in order to segregate the different periodicities present in their light curves. Figure~\ref{fig4}\,(b) and~\ref{fig5}\,(b) respectively shows the first four IMFs and their periods (P) of total intensity and Doppler velocity variation for the duration of 20\,minutes at the location B. The first four IMFs (IMF0, IMF1, IMF2, and IMF3) are observed to contain the short-period variabilities (0.2--2\,min). The successive IMFs are observed to have periodicities of more than 2\,min and hence not discussed in the present analysis. To perform a statistical study of correlation and phase-relationship between Doppler velocity and total intensity, we study 40 different light-curves (cases), each of duration 20\,minutes. These cases are selected to be located in the close neighbourhood of the power contours of 1--2\,min periodicities (red contours in the average SJ image in Figure~\ref{fig3}). The locations of the selected cases are marked in \textit{black} along the slit in the SJ image in Figure~\ref{fig3}. Few specific time-intervals are considered at these locations in order to have further study about phase-relationship between Doppler velocity and total intensity.

\begin{figure}[htbp]
	\centering
	
	\includegraphics[width=18cm]{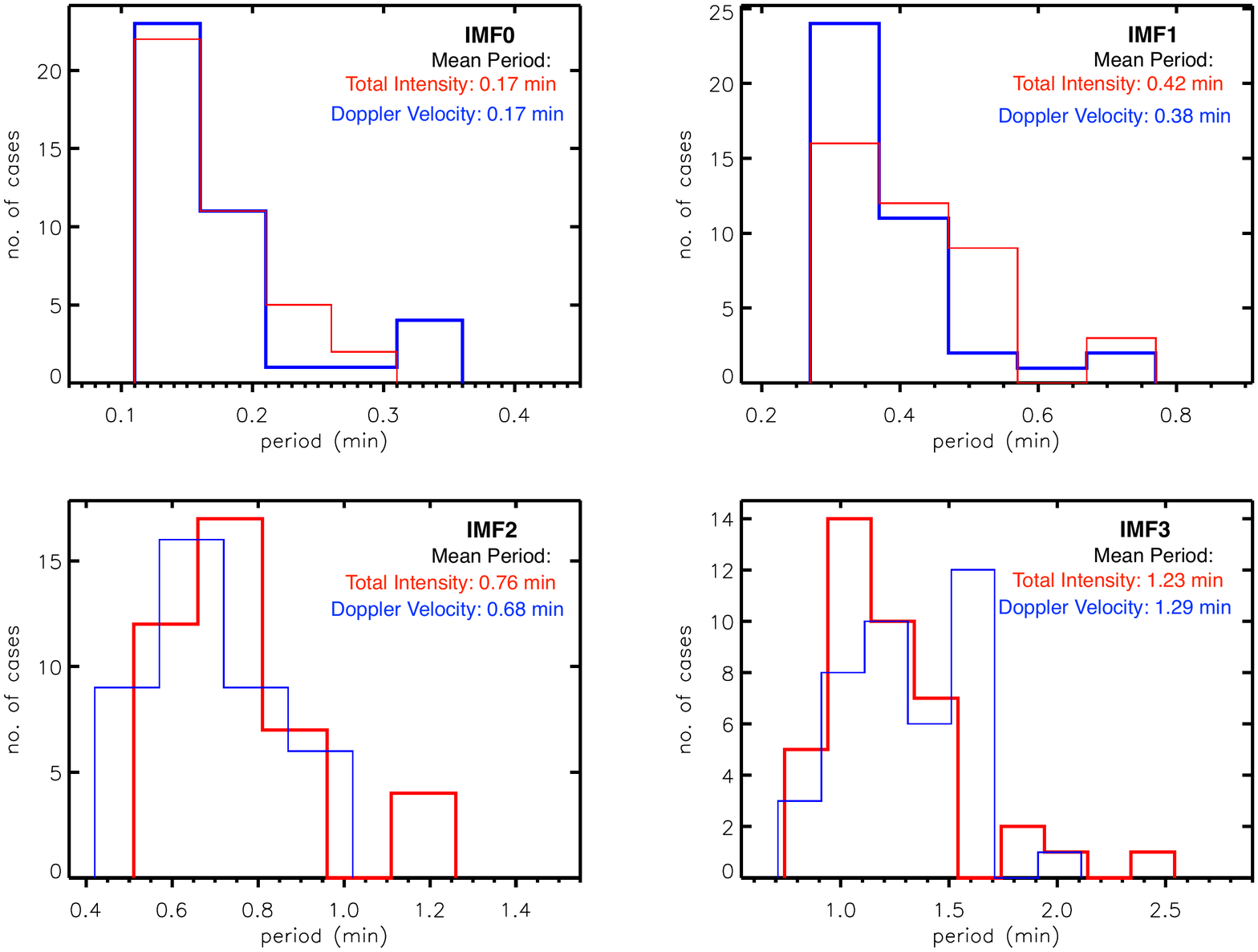}
	\caption{Histograms showing the distribution of periods of the first four IMFs of total intensity in red and Doppler velocity in blue for the 40 selected cases.}
	
	\label{fig8}
\end{figure}

\begin{figure}[htbp]
	\centering
	
	\includegraphics[width=17cm]{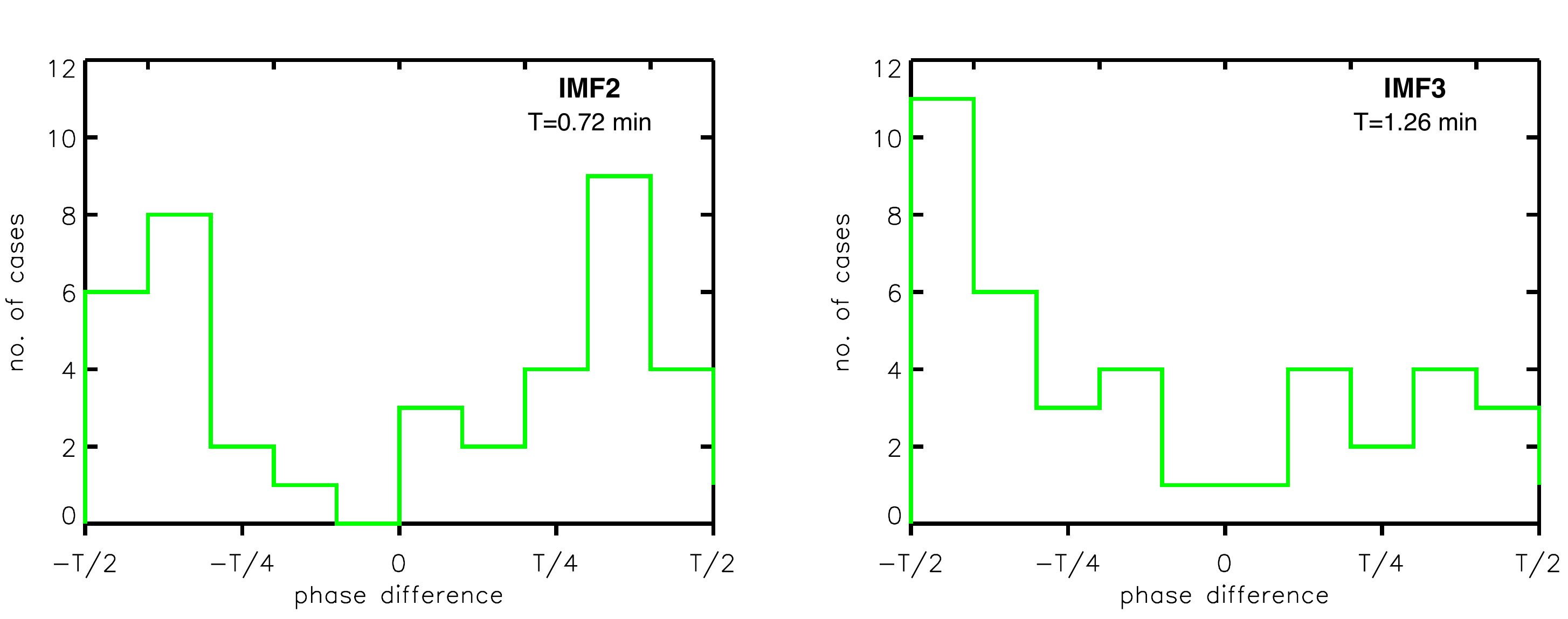}
	\caption{Histograms showing the distribution of phase-difference between total intensity and Doppler velocity for IMF2 and IMF3 for the 40 selected cases.}
	
	\label{fig9}
\end{figure}

Figure~\ref{fig8} shows the histograms of the period of oscillations for different IMFs of total intensity and Doppler velocity with the mean periods listed in the figure. As reflected by the value of mean periods, we will further regard the IMF0 to be associated with the periodicity of $\sim$0.17\,min, IMF1 with $\sim$0.40\,min, IMF2 with $\sim$0.72\,min and IMF3 with $\sim$1.26\,min. The power maps in Figure~\ref{fig7} shows the absence of significant power in the periods below 0.6\,min. Henceforth, for the further analysis about phase-relationship, we consider only the third and fourth IMFs, \textit{i.e.}, IMF2 and IMF3. The phase-relation between Doppler velocity and total intensity at the short-periodicities is studied by correlating their respective IMFs for the 40 cases.  Figure~\ref{fig9} shows the histograms of the phase difference between Doppler velocity and total intensity by considering IMF2 and IMF3. The sign convention for the values of phase-shifts considered here is such that the positive values phase-shift signifies the Doppler velocity to be leading with respect to the total intensity. The histogram of IMF2 reveal the presence of preferred phase-shifts at $\sim\pm3T/8~~(\sim\pm3\pi/4)$ where T=0.72\,min,~is the time period of oscillation. The histogram of IMF3 shows the dominant phase-shifts at $\sim-T/2~~(\sim-\pi)$ where T=1.26\,min.

The presence of a dominant phase shift of $\sim$T/2 for periodicities of $\sim$1.26\,min (IMF3) indicates the presence of reconnection events. As shown in Figure~\ref{fig6}, the increase in the intensity is accompanied by the increase in the Doppler width and decrease in Doppler velocity (blue-shifted flows, \citealp{depontieu09, depontieu10}) at many instances throughout the light-curve. Few of such instances are shown by vertical dotted lines in Figure~\ref{fig6}. In case of phase-shifts of $\sim$$\pm$T/2 or $\sim$$\pm\pi$, the reconnection process results in near simultaneous variation in the spectral parameters with the resultant mass flow projected towards the line-of-sight (blue-shifts or negative Doppler velocity). As the TR emission lines are red-shifted in general, the flows towards the line-of-sight (blue-shifted flows) will appear to decrease the Doppler speeds simultaneously with the increase in the intensity and width (phase shifts of $\sim$$\pm$T/2 or $\sim$$\pm\pi$). On the other hand, the flows away from the line-of-sight will increase the value of the Doppler speeds with an increase in the line intensity and width ($\sim$ zero phase-shift). It can be observed from Figure~\ref{fig6} and also indicated by Figure~\ref{fig9} that the red-shifted flows (cases with zero phase difference) occur less frequently compared to blue-shifted flows (cases with phase shifts of $\pm$T/2). As shown in Figure~\ref{fig6}, the instances of large amplitude fluctuations, which mostly have phase shift of $\sim$T/2 between Doppler velocity and total intensity, can be regarded as the clear signatures of quasi-periodic outflows (towards the observer) resulting from the reconnection process. The other instances of small amplitude fluctuations can be due the presence of slow magneto-acoustic waves.

\begin{figure}[htbp]
	\centering
	
	\includegraphics[width=16cm]{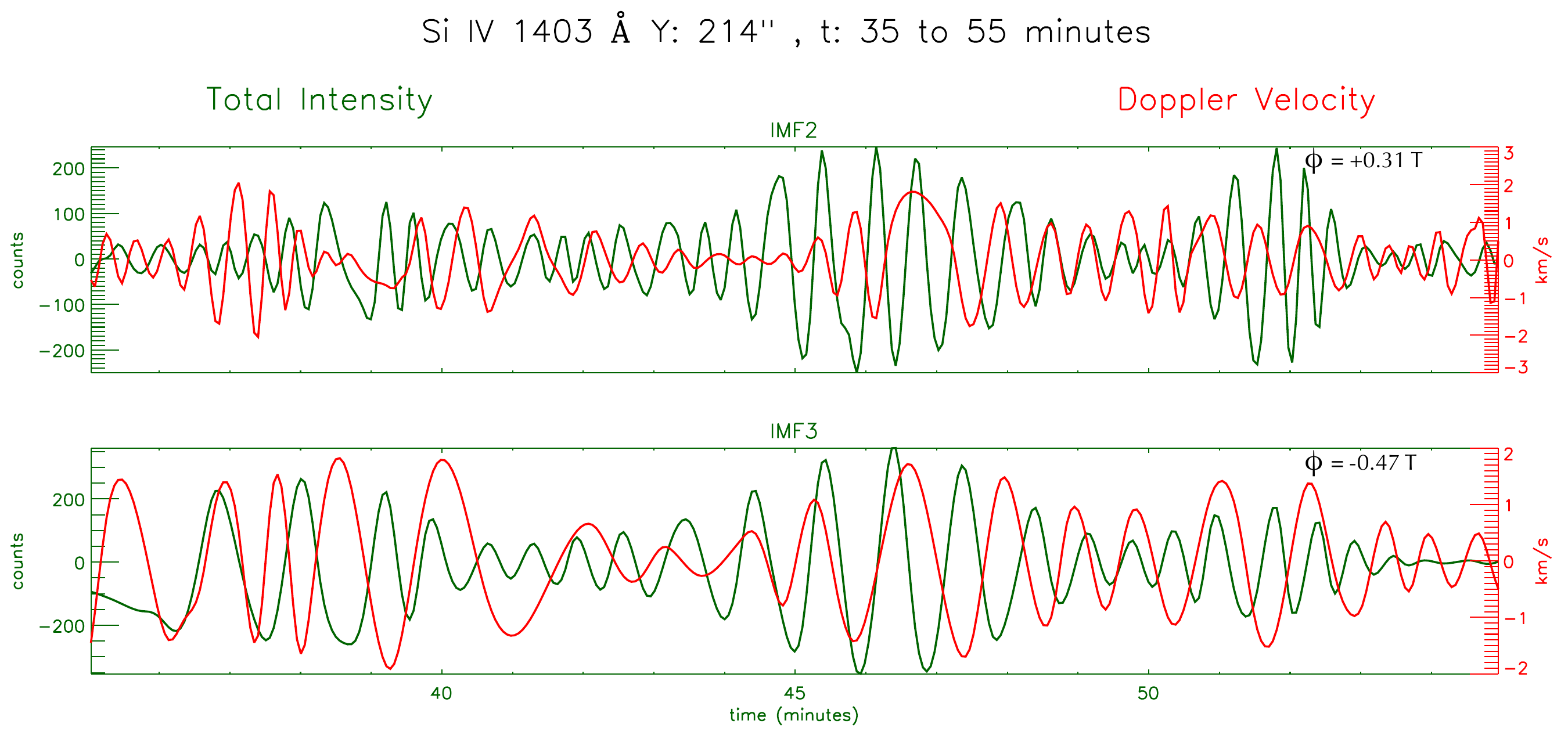}
	
	\caption{Representative examples of the IMFs at the location B along the slit, showing the respective comparisons between Doppler velocity and total intensity oscillations, and depicting that the phase-shift between them changes continuously.}
	
	\label{fig11}
\end{figure}

Very recently, \citet{hansteen14} and \citet {brooks16} have reported the presence of transition region fine loops with the aid of IRIS observations and numerical simulations. Such small scale loops with loop lengths of $\sim$1 to 2 Mm can harbour slow standing waves with periods of $\sim$1\,min in transition region. It is worth noting at this point that \citet{wang03, taroyan07, taroyan08} reported the presence of standing slow waves exclusively in hot coronal loops.  In addition, \citet{pant17} reported the existence of standing slow waves in cool coronal loops ($\sim$0.6 MK). In this work, we found evidence of the existence of slow waves in Si\,IV\,1403\,\AA~emission line whose formation temperature is $\sim$60000 K. In an ideal case, the phase-shift of $\sim$$\pm$T/4 is attributed to the presence of standing slow waves in the solar atmosphere \citep{wang03, taroyan07, taroyan08, Moreels13}. Further, it should be noted that the intensity and velocity changes phase in time due to the heating and cooling of the plasma \citep{taroyan08} and due to presence of imperfect waveguides and drivers in reality, which deviates from the theoretical considerations \citep{Keys18}. Thus the phase shift between intensity and  velocity oscillations might differ in different regions and different time as showcased in Figure~\ref{fig11}. Figure~\ref{fig11} shows the representative examples of the IMFs (IMF2 and IMF3) at the location B along the slit. The phase-shift between Doppler velocity and total intensity ($\phi$) obtained using the correlation techniques is also mentioned in the respective panels. The comparison between of the respective IMFs of intensity and Doppler velocity fluctuations clearly depicts that the phase-shift between them changes continuously throughout the entire duration. This could be due to the intermittent nature of the flows and waves that might result in departure from the theoretically expected values of the phase-shifts. Hence we conjecture that the statistically dominant phase shift of $\sim$$\pm$3T/8 for periodicities of $\sim$0.72\,min (IMF2) is due to the presence of small-scale flows along with slow standing waves in TR fine loops.
This supports both wave and reconnection like scenario to be responsible for the periodicities of 1-2\,min in moss regions, which is discussed in details in section~\ref{sec:con}.

\subsection{Density diagnostics from Si IV 1403\,\AA~ and O IV 1401\,\AA~emission lines}

\label{sec:density}

In order to obtain the information about density variations associated with the presence of waves and/or reconnection flows,  in the moss regions, we attempt to estimate density along the slit using Si\,IV\,1403\,\AA~($\lambda=1402.77$\,\AA) and O\,IV\,1401\,\AA~($\lambda=1401.16$\,\AA) spectral lines from the IRIS spectra (as suggested by~\citealp{young18a}). They introduced an empirical correction factor to normalize Si\,IV$/$O\,IV line intensity ratios. As first mentioned by~\citet{dupree72}, the observed intensities of lines from the lithium and sodium-like iso-electronic sequences are usually stronger than that expected by the emission measures from other sequences formed at the same temperature. Hence, such a correction factor is important to be applied to silicon line intensities. Table~2 of~\citet{young18a}~ gives the theoretical ratios of different lines after employing the correction factor (see QS DEM method as explained in~\citealp{young18a, young18b}). We use the Si\,IV\,(1402.77)$/$O\,IV\,(1401.16) line ratio from Table~2 of~\citet{young18a} for the estimation of electron density at a temperature of $\log~T/K=4.88$ (temperature of maximum ionization of Si\,IV).

\begin{figure}[htbp]
	\centering
	
	\includegraphics[width=17cm]{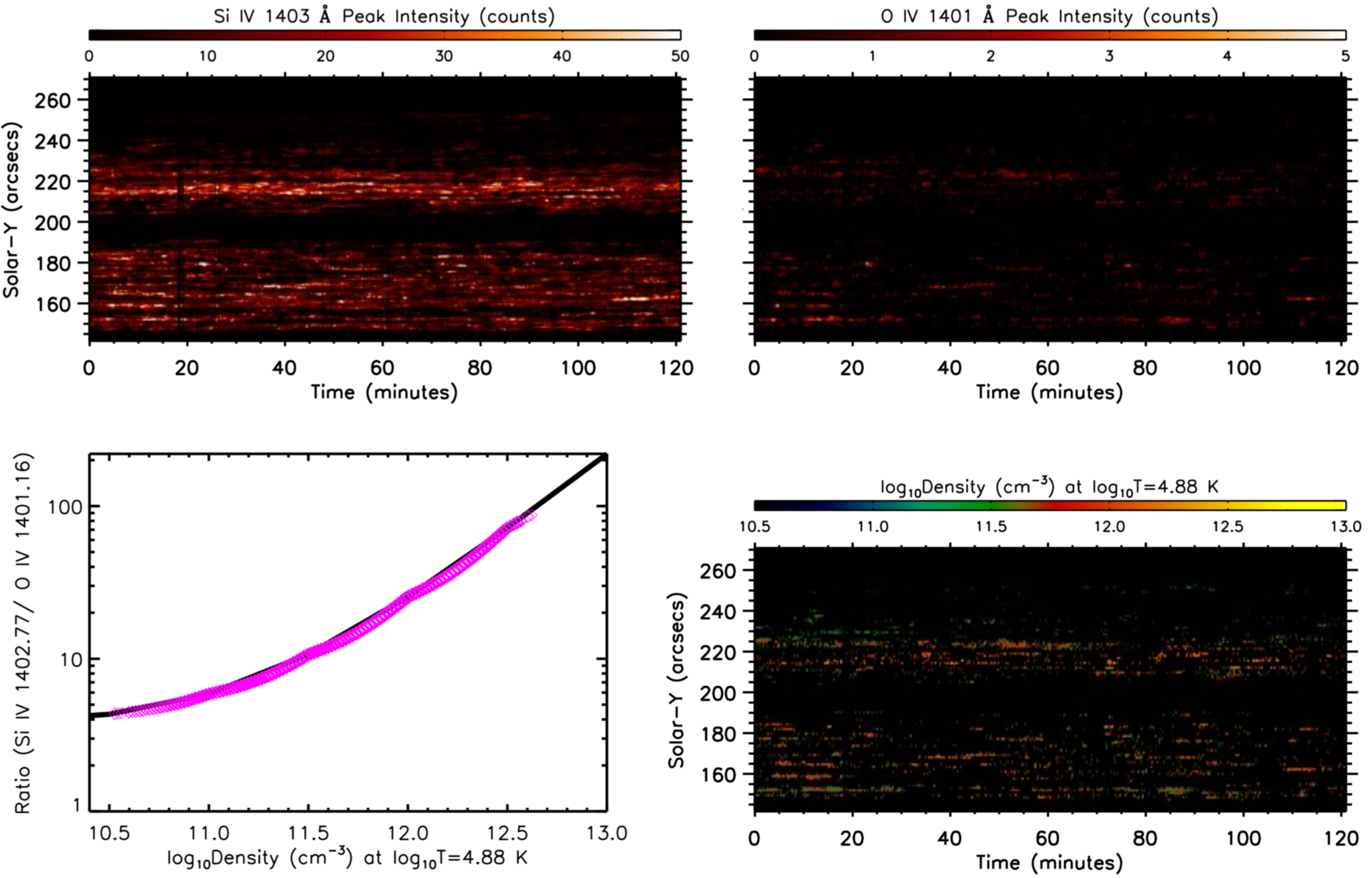}
	
	\caption{(a),(b) Time-sequence maps of peak intensity of the Si\,IV\,(1403\,\AA) and O\,IV\,(1401\,\AA) line. (c) Theoretical Si\,IV\,(1403)$/$O\,IV\,(1401) ratio–density curve in \textit{solid black} and the estimated values of the density in \textit{magenta} corresponding to the observed ratios. (d) Time-sequence map of the estimated density.}
	
	\label{fig10}
\end{figure}

As the O\,IV\,1401\,\AA~line is very weak in IRIS spectra, the spectra is averaged over 7 pixels along the slit. In such averaging, for instance, the data value of the first 7 spatial pixels are replaced by their average value, the next 7 pixels are replaced by their respective average data-value, and so on. Similarly, time-averaging is also performed by considering 4 time steps along the temporal axis. In order to improve S/N,  such averaging is performed only over  O\,IV\,1401\,\AA \,spectra~as Si\,IV\,1403\,\AA~spectra contains significantly good signal. Figure~\ref{fig10}\,(a)\,and\,(b) shows the time-sequence maps of peak intensity along the slit for the Si\,IV\,1403\,\AA~and O\,IV\,1401\,\AA~line-profiles. A comparison between the two maps clearly shows that despite averaging the spectra (as explained above), we are able to obtain good S/N only for very few isolated O\,IV\,1401\,\AA~line-profiles in order to perform a reliable Gaussian fit, hence the peak intensity values for the O\,IV\,1401\,\AA~line are shown only for those isolated few pixels.

Figure~\ref{fig10}(c)  shows the theoretical Si\,IV\,(1403)$/$O\,IV\,(1401) ratio-density curve~\citep{young18a} in \textit{solid black} and the estimated density values are over-plotted in  \textit{magenta}. The density time-sequence map is also showcased in Figure~\ref{fig10}(d). Note that we could estimate the density only at very few instances of some of the locations, as limited by the poor signal in O\,IV\,1401\,\AA~spectra. It can be observed in Figure~\ref{fig10}(d) that we cannot find considerable examples of continuous density signal along time for some significant amount of duration over the entire observation. It is completely  unreliable to perform any time series analysis over such light-curves. It appears that there are definite changes in the density but to relate those changes with intensity and other line parameters for identification of the wave mode is beyond the quality of the current observations. Thus, we are still unable to obtain any results related to  density oscillations with the present data.

\section{Conclusions}
\label{sec:con}

In the present article, we study high-frequency dynamics of active region moss by using high spatially and spectrally resolved observations of IRIS, with the fast cadence of 13\,seconds for imaging and 3.3\,seconds for spectral data. The techniques of wavelet and EMD analysis are employed in conjunction to explore the characteristics of the high-frequency oscillations. We have observed the persistent presence of periodicities in the 1--2\,min range in the Si\,IV\,1400\,\AA~SJ intensity as well as in different spectral parameters (total intensity, peak intensity, Doppler velocity, and Doppler width) derived from the Si\,IV\,1403\,\AA~emission line. The power maps deduced from the SJ intensity variations show the concentration of power in short-periodicities generally in the bright regions of the moss. This result is in agreement with the study of \citet{pant15}, where the authors reported high-frequency quasi-periodic oscillations concentrated over localised regions in the active region moss. However, no attempts were made to understand the nature of variability due to the lack of spectral data. That study was performed using the 193\,\AA~passband of Hi-C which is sensitive to coronal temperatures. In this work, we find similar signatures in TR. Additionally, the power maps of the spectral parameters also reveal the predominance of significant power in the 1--2\,min period range.

We study the phase difference between Doppler velocities and total intensity. Our study supports both wave and reconnection like scenario to be responsible for the periodicities of 1-2\,min in moss regions. Studying the phase relationships, we can conclude that the periodicity of 1.26\,min with dominant phase shifts of $\sim-T/2~~(\sim-\pi)$ is predominantly due to the outflows resulting from the reconnection process. On the other hand, the periodicity of 0.72\,min with dominant phase shifts of $\sim\pm3T/8~~(\sim\pm3\pi/4)$  can be regarded as the collective signatures of the small-scale flows and slow standing modes existing within the transition region fine loops of lengths 1 to 2 Mm. Hence qualitatively, we can conjecture that the high-frequency oscillations of $\sim$1\,min, observed in the bright moss regions are possibly due to the combination of slow magneto-acoustic waves and reconnection events. As explained in section~\ref{sec:density}, we cannot obtain any reliable results from the density variations, although we are able to estimate the average density of the moss regions but to reliably  study the density variation much better quality of data is required. The high-frequency oscillations in the moss regions can be due to compressive waves. The key to distinguish between the different modes conclusively is to study the density variations which is not possible with present data because of low data-counts present in the O\,IV\,1410\,\AA~emission line. Some new instruments, with better sensitivity in the FUV wavelengths, especially in the density sensitive lines, may provide new insight and will enable us to specifically detect the particular wave modes responsible for such oscillations.

\section*{Author Contributions}
VP identified the IRIS data. VP and DB planned the study. NN performed all analysis and wrote the manuscript.  VP, DB, and TVD helped in analysing the results. All authors participated in the discussion. 

\section*{Funding}
NN is supported by the Senior Research Fellowship scheme of the Council of Scientific and Industrial Research (CSIR) under the Human Resource Development Group (HRDG), India. TVD and VP are supported by the GOA-2015-014 (KU~Leuven) and the European Research Council (ERC) under the European Union's Horizon 2020 research and innovation programme (grant agreement No 724326).

\section*{Acknowledgments}
Authors acknowledge IRIS team for the publicly available data used in this paper. IRIS is a NASA Small Explorer mission developed and operated by the Lockheed Martin Solar and Astrophysics Laboratory (LMSAL), with mission operations executed at the NASA Ames Research Center and major contributions to downlink communications funded by the Norwegian Space Center (Norway) through a European Space Agency PRODEX contract.

\section*{Supplemental Data}
 \href{http://home.frontiersin.org/about/author-guidelines#SupplementaryMaterial}{Supplementary Material} should be uploaded separately on submission, if there are Supplementary Figures, please include the caption in the same file as the figure. LaTeX Supplementary Material templates can be found in the Frontiers LaTeX folder.

\section*{Data Availability Statement}
The datasets analysed in this study can be found at the IRIS website maintained by LMSAL (\hyperlink{http://iris.lmsal.com}{http://iris.lmsal.com}).

\bibliographystyle{frontiersinSCNS_ENG_HUMS} 


\end{document}